\documentclass[10pt,table,xcdraw,conference]{IEEEtran}

\usepackage{graphicx}
\usepackage{microtype}
\usepackage{hyperref}
\usepackage{amsmath}
\usepackage{amssymb}

\usepackage{multirow}
\usepackage{tabularx,adjustbox,booktabs}
\usepackage{siunitx}

\usepackage{array}
\newcolumntype{$}{>{\global\let\currentrowstyle\relax}}
\newcolumntype{^}{>{\currentrowstyle}}
\newcommand{\rowstyle}[1]{\gdef\currentrowstyle{#1}%
    #1\ignorespaces
}

\usepackage{mathtools}
\usepackage[]{xcolor}

\usepackage[skip=2pt]{caption} 
\usepackage{tikz}
\usetikzlibrary{shapes,arrows}

\RequirePackage{titlesec}
\titlespacing*{\section}{2pt}{1.5ex}{1.5ex}
\titlespacing*{\subsection}{2pt}{1ex}{1ex}

\newcommand\restr[2]{{
  \left.\kern-\nulldelimiterspace 
  #1 
  \vphantom{\big|} 
  \right|_{#2} 
  }}
 
\newcommand{\changed}[1]{\textcolor{black}{#1}}  

\usepackage[numbers,sort&compress]{natbib}
\usepackage{listings}

\newcommand{\spl}{software product line}

\newcommand{\eo}{edit operation}

\newcommand{\figref}[1]{Figure~\ref{#1}}
\newcommand{\secref}[1]{Section~\ref{#1}}

\usepackage{todonotes}

\usepackage{amsthm}

\theoremstyle{definition}

\begin{document}

\title{Learning Domain-Specific Edit Operations from Model Repositories with Frequent~Subgraph~Mining}


\author{\IEEEauthorblockN{
Christof Tinnes\IEEEauthorrefmark{2},
Timo Kehrer\IEEEauthorrefmark{1},
Mitchell Joblin\IEEEauthorrefmark{2},
Uwe Hohenstein\IEEEauthorrefmark{2},
Andreas Biesdorf\IEEEauthorrefmark{2},
Sven Apel\IEEEauthorrefmark{3}}
\IEEEauthorblockA{\IEEEauthorrefmark{2}Siemens AG - Corporate Technology, 81739 M{\"u}nchen, Germany \\
\{christof.tinnes,mitchell.joblin,uwe.hohenstein\}@siemens.com,
\IEEEauthorrefmark{2}Humboldt-Universität zu Berlin, 12489 Berlin-Adlershof, Germany \\
\{timo.kehrer\}@informatik.hu-berlin.de,
\IEEEauthorrefmark{2}Saarland University, 66123 Saarbrücken, Germany \\
\{apel\}@cs.uni-saarland.de}
}

%



\maketitle              
\newcommand{\code}[1]{\small {\sl{#1}}}

\begin{abstract}
Model transformations play a fundamental role in model-driven software development. They can be used to solve or support central tasks, such as creating  models, handling model co-evolution, and model merging. In the past, various (semi-)automatic approaches have been proposed to derive model transformations from meta-models or from examples. These approaches require time-consuming handcrafting or recording of concrete examples, or they are unable to derive complex transformations. We propose a novel \emph{unsupervised} approach, called \textsc{Ockham}, which is able to learn edit operations from model histories in model repositories. \textsc{Ockham} is based on the idea that meaningful \eo{s} will be the ones that \emph{compress} the model differences. We evaluate our approach in two controlled experiments and one real-world case study of a large-scale industrial model-driven architecture project in the railway domain. We find that our approach is able to discover frequent edit operations that have actually been applied. Furthermore, \textsc{Ockham} is able to extract edit operations in an industrial setting that are meaningful to practitioners.
\end{abstract}

\section{Introduction}
Software and systems become increasingly complex. Various languages, methodologies, and paradigms have been developed to tackle this complexity.
One widely-used methodology is Model-Driven Engineering (MDE) \cite{RodriguesDaSilva2015}, which uses models as first class entities and facilitates generating documentation and (parts of the) source code from these models. Usually, Domain-Specific Modeling Languages are used and tailored to the specific needs of a domain. This reduces the cognitive distance between the domain and the used language. A key ingredient of many tasks and activities in MDE are model transformations \cite{Sendall2003}. 

In this paper, we focus on \eo{s} as an important subclass of model transformations. An edit operation is an in-place 
model transformation and usually represents regular evolution \cite{visser2007model} of the models. For example, when moving a method from one class to another in a class diagram, also a sequence diagram that uses the method in message calls between object lifelines needs to be adjusted (e.g., by changing the receiver of a message accordingly). To perform this in a single edit step, one can create an \eo{} that executes the entire change, including the class and sequence diagram changes.
Some tasks can even be completely automatized and reduced to the definition of \eo{s}. Edit operations are used for model repair, quick-fix generation, auto completion~\cite{ohrndorf2021history, hegedus2011quick, Kogel2016}, model editors \cite{taentzer2007generating, ehrig2005generation},  operation-based merging \cite{koegel2009operation}, model refactoring \cite{Mokaddem2018, Arendt2013},  model optimization \cite{burdusel2018mdeoptimiser}, meta-model evolution and model co-evolution \cite{rose2014graph, arendt2010henshin, herrmannsdoerfer2010extensive}, artifact co-evolution in general \cite{getir2018supporting, kolovos2010taming}, semantic lifting of model differences \cite{Kehrer2011, kehrer2012understanding, ben2012search, langer2013posteriori, khelladi2016detecting}, model generation \cite{pietsch2011generating}, and many more.

In general, there are two main problems involved in the specification of model transformations than can be used as \eo{s}.
Firstly, creating the necessary transformations for the task and the Domain-Specific Modeling Languages at hand using a dedicated transformation language requires a deep knowledge of the Domain-Specific Modeling Language's meta-model and the underlying paradigm of the transformation language. It might even be necessary to define project-specific \eo{s}, which causes a large overhead for many projects or tool providers \cite{Kehrer2017, Mokaddem2018, Kappel2012}.
Secondly, for some tasks, the domain-specific transformations are only a form of tacit knowledge \cite{Polanyi1958}, and it will be hard for domain experts to externalize this knowledge.

Because, on the one hand, model transformations play such a central role in MDE, but, on the other hand, it's not easy to specify them, attempts have been made to support their manual creation or even (semi\nobreakdash-)automated generation.
As for manual support, visual assistance tools \cite{Avazpour2015} and transformation languages derived from a modeling language's concrete syntax \cite{acrectoaie2018vmtl, holldobler2015systematically}  have been proposed to release domain experts from the need of stepping into the details of meta-models and model transformation languages. However, they still need to deal with the syntax and semantics of certain change annotations, and edit operations must be specified in a manual fashion.
To that end, generating edit operations automatically from a given meta-model has been proposed \cite{Kehrer2016, mazanek2009generating}. However, besides elementary consistency constraints and basic well-formedness rules, meta-models do not convey any domain-specific information on how models are edited. Thus, the generation of edit operations from meta-models is limited to rather primitive operations as a matter of fact. 
Following the idea of model transformation by-example (MTBE) \cite{Langer2009, Gray2011, Kappel2012}, initial sketches of more complex and domain-specific edit operations can be specified using standard model editors as a macro recorder. However, these sketches require manual post-processing to be turned into general specifications, mainly because an initial specification is derived from only a single transformation example.
Some MTBE approaches \cite{Kehrer2017, Mokaddem2018} aim at getting rid of this limitation by using a set of transformation examples as input which are then generalized into a model transformation rule. Still, this is a supervised approach which requires sets of dedicated transformation examples that need to be defined by domain experts in a manual fashion. As discussed by Kehrer et al.~\cite{Kehrer2017}, a particular challenge is that domain experts need to have, at least, some basic knowledge on the internal processing of the MTBE tool in order to come up with a reasonable set of examples. Moreover, if only a few examples are used as input for learning, Mokaddem et al. \cite{Mokaddem2018} discuss how critical it is to carefully select and design these examples. 



To address these limitations of existing approaches, we propose a novel unsupervised approach, \textsc{Ockham}, for mining \eo{s} from existing models in a model repository, which is typically available in large-scale modeling projects (cf.\ \secref{sec:case-study}). \textsc{Ockham} is based on an Occam's razor argument, that is, the ``useful'' \eo{s} are the ones that ``compress'' the model repository. In a first step, \textsc{Ockham} discovers frequent change patterns using frequent subgraph mining on a labeled graph representation of model differences. It then uses a compression metric to filter and rank these patterns. We evaluate our approach using two experiments with simulated data and one real-world large-scale industrial case study from the railway domain. In the simulated cases, we can show that \textsc{Ockham} is able to discover the \eo{s} that have been actually applied in the simulation, even when we apply some ``perturbation''. In the real-world case study, we find that our approach is able to scale to real-world model repositories and to derive \eo{s}. We evaluated \textsc{Ockham} by comparing the results to randomly generated \eo{s} in five interviews with practitioners of the product line. We find that the \eo{s} represent typical edit scenarios and are meaningful to the practitioners.

In a summary, we make the following contributions:
\begin{itemize}
    \item We propose an unsupervised approach based on frequent subgraph mining to derive \eo{s} out of model repositories, without requiring any further information (e.g., labeling).
    \item We evaluate our approach empirically based on two controlled simulated experiments and show that the approach is able to discover the actually applied \eo{s}.
    \item We evaluate the approach using an interview with five experienced system engineers and architects from a real-world industrial setting in the railway domain with more than 200 engineers, 300GB of artifacts and more than 6 years of modeling history. We show that our approach is able to detect meaningful \eo{s} in this industrial setting and to scale to real-world repositories.
\end{itemize}

\section{\changed{Motivation: An Industrial Scenario}}\label{sec:case-study}
Our initial motivation to automatically mine \eo{s} from model repositories arose from a long-term collaboration with practitioners from a large-scale industrial model-driven \spl{} in the railway domain. The modeling is done in \textsc{MagicDraw} \cite{MagicDraw} using SysML, and there is an export to the Eclipse Modeling Framework (EMF) which focuses on the SysML parts required for subsequent MDE activities (e.g., code generation). Modeling tools such as \textsc{MagicDraw} 
come with support for model versioning. In our case, the models are versioned in the MagicDraw Teamwork Server. We therefore have access to a large number of models and change scenarios. \par 
During discussing major challenges with the engineers of the product line, we observed that some model changes appear very often together in this repository. For example, when the architect creates an interface between two components, s/he will usually add some \emph{Ports} to \emph{Components} and connect them via the \emph{ConnectorEnds} of a \emph{Connector}. Expressed in terms of the meta-model, there are 17 changes to add this interface. We are therefore interested if we can automatically detect these patterns in the model repository. More generally, our approach, \textsc{Ockham}, is based on the assumption that it should be possible to derive ``meaningful'' patterns from the repositories.\par

These patterns could then be used for many applications~\changed{\cite{ohrndorf2021history, Kogel2016, taentzer2007generating, Arendt2013,  getir2018supporting, kehrer2012understanding, ben2012search, langer2013posteriori, khelladi2016detecting}}. 
In our case study, the models have become huge over time (approx. 1.2 million elements split into 100 submodels) and also model differences between different products have become huge (up to 190.000 changes in a single submodel). The analysis of these differences, e.g., for quality assurance of the models, or domain analysis has become time-consuming. To speed-up the analysis of the model differences, it would be desirable to reduce the ``perceived'' size of the model difference by grouping fine-grained differences to higher-level, more coarse-grained and more meaningful changes. For this semantic lifting of model differences, the approach by Kehrer et al.~\cite{Kehrer2011}, which uses a set of \eo{s} as configuration input, can be used. These large model differences have actually been our main motivation to investigate how we can derive the required \eo{s} (semi-)automatically. \par
We will use the data from this \changed{real-world project} to evaluate \textsc{Ockham} in \secref{sec:evaluation}.

\section{Background}\label{sec:background}
In this section, we provide basic definitions that are important to understand our approach presented in \secref{sec:appraoch}.

\subsection{Graph theory}
\begin{figure}[ht]
			\centering
			\includegraphics[width=\columnwidth, clip]{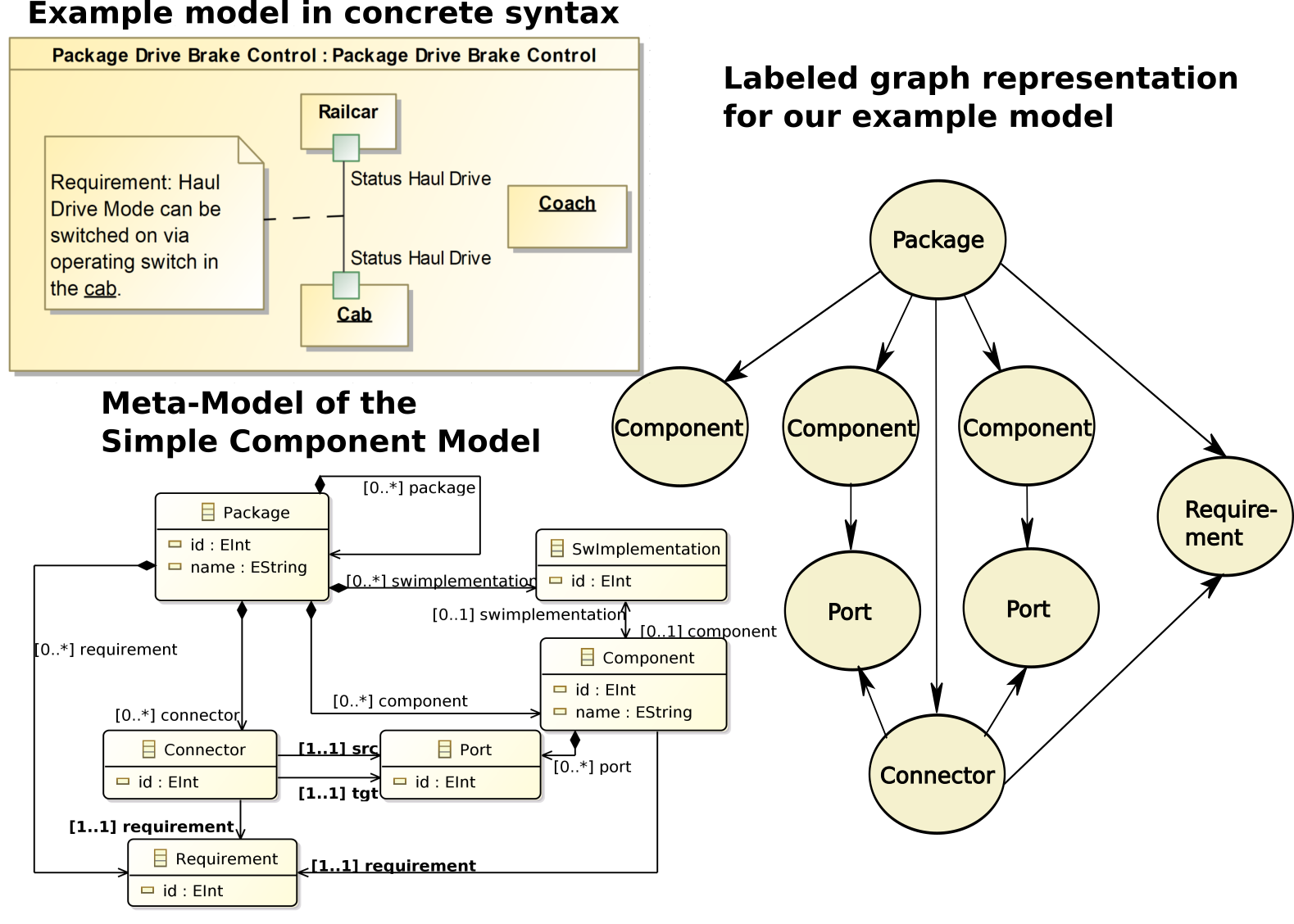}
			\caption{We consider models as labeled graphs, where labels represent types of nodes and edges defined by a meta model. For the sake of brevity, types of edges are omitted in the figure.}
			\label{fig:difference-graph}
			\vspace{-1em}
\end{figure}

As usual in MDE, we assume that a meta-model specifies the abstract syntax and static semantics of a modeling language. We conceptually consider 
a model as a typed graph (aka. abstract syntax graph), in which the types of nodes and edges are drawn from the meta-model. \figref{fig:difference-graph} illustrates how a simplified excerpt from an architectural model of our case study from \secref{sec:case-study} in concrete syntax is represented in its abstract syntax, typed over the given meta-model. 

Since we further assume models to be correctly typed, in our notion of a graph used throughout the paper, we abstain from a formal definition of typing using type graphs and type morphisms \cite{Biermann2012}. Instead, to keep our basic definitions as simple as possible, we work with a variant of labeled graphs where a fixed label alphabet represents node and edge type definitions of a meta-model.
Given a label alphabet $L$, a \textit{labeled directed graph} $G$ is a tuple $(V,E,\lambda)$, where $V$ is a finite set of nodes, $E$ is a subset of $V \times V$, called the edge set, and $\lambda: V \cup E \to L$ is the labeling function, which assigns a label to nodes and edges. 
If we are only interested in the structure of a graph and typing is irrelevant, we will omit the labeling and only refer to the graph as $G = (V,E)$.


Given two graphs $G = (V,E,\lambda)$ and $G' = (V',E',\lambda')$, $G'$ is called a \emph{subgraph} of $G$, written $G' \subseteq G$, if $V' \subseteq V$, $E' \subseteq E$, and 
$\lambda(x) = \lambda'(x)$ for each $x \in V' \cup E'$.
\changed{
A \textit{(weakly) connected component} (component, for short) $C=(V_C, E_C) \subseteq G$ is an induced subgraph of $G$ in which every two vertices are connected by a \textit{path}, that is,  
${\forall u,v \in V_C} : {\exists n\in\mathbb{N}} \text{ s. t. } {\big\lbrace (v, v_1), (v_1, v_2), \dots ,(v_n, u)\big\rbrace} \subseteq E_C \ \cup \tilde E_C$, where $\tilde E_C$ is the set of all reversed edges, that is, $(u, v) \in E_C$ becomes $(v, u) \in \tilde E_C$.
}
\subsection{Frequent Subgraph Mining}
We will use frequent subgraph mining as the main ingredient for \textsc{Ockham}. We distinguish between \textit{graph-transaction-based frequent subgraph mining} and \textit{single-graph-based frequent subgraph mining}.
Graph-transaction-based frequent subgraph mining uses a collection (aka. \emph{database}) of graphs, while single-graph-based frequent subgraph mining looks for subgraphs of a single graph. 
We are considering graph-transaction-based frequent subgraph mining in this work. 
A subgraph mining algorithm typically takes a database of graphs and a threshold $t$ as input. It then outputs all the subgraphs with, at least, $t$ occurrences in the database. 
An overview of the frequent subgraph mining algorithms can be found in the literature \cite{Jiang2008}. \changed{
A general introduction to graph mining is given by Cook and Holder \cite{cook2006mining}, who also proposed a compression-based subgraph miner called \textsc{Subdue}~\cite{Ketkar2005}.  \textsc{Subdue} has also been one of our main inspirations for a compression-based approach. 
}\textsc{Ockham} is based on \textsc{Gaston} \cite{Nijssen2005}, which mines frequent subgraphs by first focusing on frequent paths, then extending to frequent trees, and finally extending the trees to cyclic graphs.

\subsection{Model Transformations and Edit Operations}\label{sec:editoperations}
The goal of \textsc{Ockham} is to learn domain-specific \textit{\eo{s}} from model histories. In general, \eo{s} can be informally understood as editing commands which can be applied to modify a given model. In turn, a difference between two model versions can be described as a (partially) ordered set of applications of \eo{s}, transforming one model version into the other. Comparing two models can thus be understood as determining the \eo{} applications that transform one model into the other.
A major class of \eo{s} are model refactorings, which induce syntactical changes without changing a models' semantics. Other classes of \eo{}s are recurring bug fixes and evolutionary changes (i.e., adding new functionality). 

In the classification given by Visser et al. \cite{visser2007model}, \eo{s} can describe regular evolution \cite{visser2007model}, that is, ``the modeling language is used to make changes'', but are not meant to describe meta-model evolution, platform evolution or abstraction evolution. More technically, in Mens et al.'s taxonomy~\cite{Mens2006}, \eo{s} can be classified as endogenous (i.e., source and target meta-model are equal), in-place (i.e., source and target model are equal) model transformations. 
For the purpose of this paper, we define an \eo{} as an in-place model transformation which represents regular model evolution.

The model transformation tool \textsc{Henshin}~\cite{arendt2010henshin} supports the specification of in-place model transformations in a declarative manner. It is based on graph transformation concepts~\cite{Ehrig2004}, and provides a visual language for the definition of transformation rules, which is used, e.g., in the last step of \figref{fig:approach}. Roughly speaking, transformation rules specify graph patterns to be found and created or deleted.

\section{Approach}\label{sec:appraoch}
We address the problem of automatically identifying \eo{s} from a graph mining perspective. 
As already discussed in \secref{sec:background}, we will work with labeled graphs instead of typed graphs. There are some limitations related to this decision, which are discussed in \secref{sec:limitations}.

\begin{figure*}[ht]
			\centering
			\includegraphics[width=2\columnwidth]{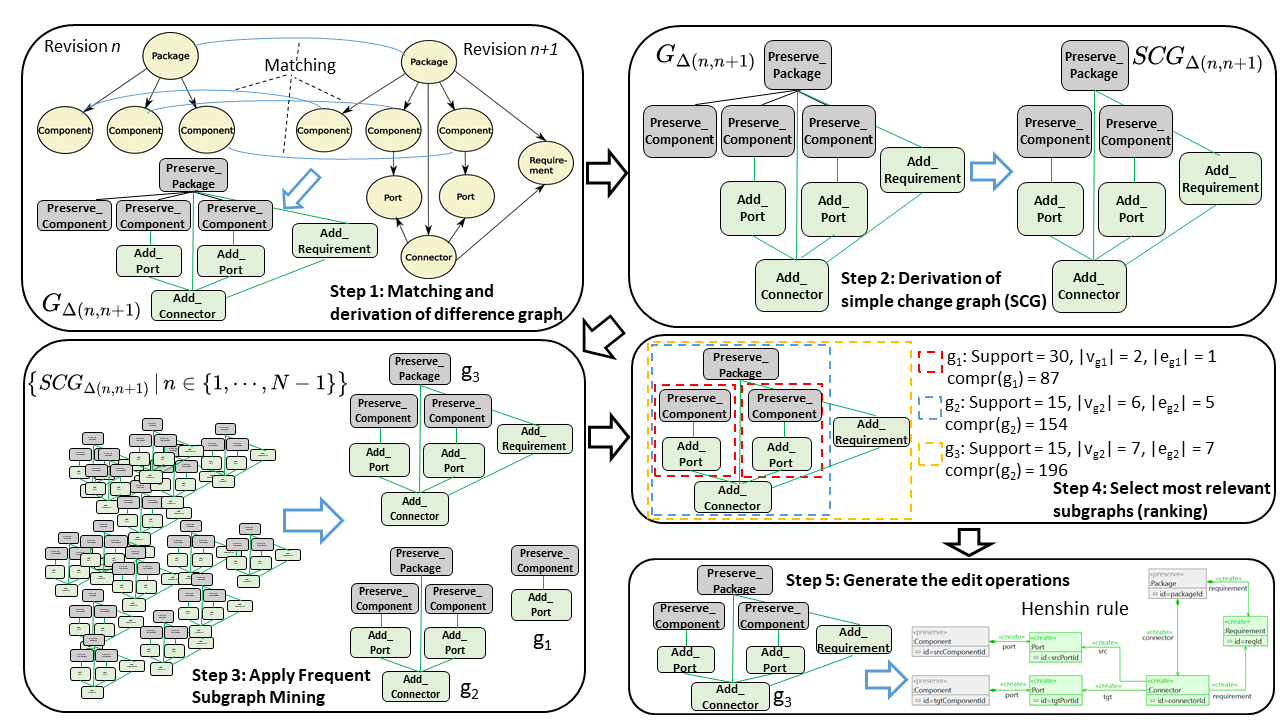}
			\caption{The 5 step process for mining \eo{s}.}
			\label{fig:approach}
			\vspace{-1em}
\end{figure*}

\textsc{Ockham} consists of the five steps illustrated with a running example in \figref{fig:approach} and outlined below.
Our main technical contributions are Step 2 and Step 4. For Step~1, Step~3, and Step~5 we apply existing tooling: \textsc{SiDiff}, \textsc{Gaston}, and \textsc{Henshin}.


\textbf{Step 1: Compute Structural Model Differences}: 
To learn a set of edit operations in an \emph{unsupervised} manner,
\textsc{Ockham} analyzes model changes which can be observed in a model's development history. In this first step, for every pair of successive model versions  $n$ and $n+1$ in a given model history, we calculate a {\em structural model difference} $\Delta(n,n+1)$ to capture these changes. Since we do not assume any information (e.g., persistent change logs) to be maintained by a model repository, we use a state-based approach to calculate a structural difference $\Delta(n,n+1)$, which proceeds in two steps~\cite{Kehrer2015}. First, the corresponding model elements in the model graphs $G_{n}$ and $G_{n+1}$  are determined using a model matcher~\cite{kolovos2009different}. Second, the structural changes are derived from these correspondences: All the elements in $G_{n}$ that do not have a corresponding partner in $G_{n+1}$ are considered to be deleted, whereas, vice versa, all the elements in $G_{n+1}$ that do not have a corresponding partner in $G_{n}$ are considered to be created.

For further processing in subsequent steps, we represent a structural difference $\Delta(n,n+1)$ in a graph-based manner, referred to as {\em difference graph}~\cite{ohrndorf2021history}. A difference graph $G_{\Delta(n,n+1)}$ is constructed as a unified graph over $G_{n}$ and $G_{n+1}$. That is, corresponding elements being preserved by an evolution step from version $n$ to $n+1$ appear only once in $G_{\Delta(n,n+1)}$  (indicated by the label prefix ``preserved\_''), while all other elements that are unique to model $G_{n}$ and $G_{n+1}$ are marked as deleted and created, respectively (indicated by the label prefixes ``delete\_'' and ``create\_'').

To give an illustration, assume that the architectural model shown in \figref{fig:difference-graph} is the revised version $n+1$ of a version $n$ by adding the ports along with the connector and its associated requirement. \figref{fig:approach} illustrates a matching of the abstract syntax graphs of the model versions $n$ and $n+1$. For the sake of brevity, only correspondences between nodes in $G_{n}$ and $G_{n+1}$ are shown in the figure, while two edges are corresponding when their source and target nodes are in a correspondence relationship. The derived difference graph $G_{\Delta(n,n+1)}$ is illustrated in \figref{fig:approach}. For example, the corresponding nodes of type {\sf Component} occur only once in $G_{\Delta(n,n+1)}$, and the nodes of type {\sf Port} are indicated as being created in version $n+1$.
Our implementation is based on the Eclipse Modeling Framework. We use the tool \textsc{SiDiff} \cite{schmidt2008constructing} to compute structural model differences. \changed{Our requirements on the model differencing tool are support for EMF, the option to implement a custom matcher, and an approach to semantically lift model differences based on a set of given \eo{s}. Modeling tools such as \textsc{MagicDraw} usually provide IDs for every model element, which can be used by a custom matcher to calculate matches based on existing IDs. We intend to use the semantic lifting approach for the compression of differences in the project from Sec. \ref{sec:case-study}. Other tools such as \textsc{EMFCompare} could also be used for the computation of model differences and there are no other criteria to favour one over the other. An overview of the different matching options is given by Kolovos et al.~\cite{kolovos2009different}; a survey of model comparison approaches is given by Stephan and Cordy~\cite{stephan2013survey}.}

\textbf{Step 2: Derive Simple Change Graphs}:
Real-world models maintained in a model repository, such as the architectural models in our case study, can get huge. It is certainly fair to say that, compared to a model's overall size, only a small number of model elements is actually subject to change in a typical evolution step. Thus, in the difference graphs obtained in the first step, the majority of difference graph elements represent model elements that are simply preserved. To that end, before we continue with the frequent subgraph mining in step 3, in step 2, difference graphs are reduced to {\em simple change graphs} (SCGs) based on the principle of \emph{locality relaxation} that only changes that are ``close'' to each other can result from the application of a single \eo{}. We discuss the implications of this principle in \secref{sec:limitations}. By ``close'', we mean that the respective difference graph elements representing a change must be directly connected (i.e., not only through a path of preserved elements). Conversely, this means that changes being represented by elements that are part of different connected components of a simple change graph are independent of each other (i.e., they are assumed to result from different edit operation applications).

More formally, given a difference graph $G_{\Delta(n,n+1)}$, a simple change graph $SCG_{\Delta(n,n+1)} \subseteq G_{\Delta(n,n+1)}$ is derived from $G_{\Delta(n,n+1)}$ in two steps. First, we select all the elements in $G_{\Delta(n,n+1)}$ representing a change (i.e., nodes and edges that are labeled as ``delete\_*'' and ``create\_*'', respectively). In general, this selection does not yield a graph, but just a graph fragment $F \subseteq G_{\Delta(n,n+1)}$, which may contain dangling edges when the source or target node of a changed edge is a preserved node not included in $F$. In a second step, these preserved nodes are also selected to be included in the simple change graph. Formally, the simple change graph is constructed as the boundary graph of $F$, which is the smallest graph $SCG_{\Delta(n,n+1)} \subseteq G_{\Delta(n,n+1)}$ completing $F$ to a graph~\cite{Kehrer2015}. 
The derivation of a simple change graph from a given difference graph is illustrated in the second step of \figref{fig:approach}. 
In this example, the simple change graph comprises only a single connected component. In a realistic setting, however, a simple change graph typically comprises a larger set of connected components, like the one illustrated in step 3 of \figref{fig:approach}.

\textbf{Step 3: Apply Frequent Connected Subgraph Mining}:
When we apply the first two steps to a model history, we obtain a set of simple change graphs $\big\lbrace SCG_{\Delta(n,n+1)} \,|\, n \in \lbrace 1,\dots ,N-1\rbrace\big\rbrace$, where N is the number of revisions in the repository. In this set, we want to identify recurring patterns and therefore find some frequent connected subgraphs. A small support threshold might lead to a huge number of frequent subgraphs. This not only causes large computational effort but also makes it difficult to find the relevant subgraphs. As it would be infeasible to recompute the threshold manually for every dataset, we pre-compute it by running an approximate frequent \textit{subtree miner} for different thresholds up to some fixed size of the frequent subtrees. We fix the range of frequent trees and adjust the threshold accordingly. Alternatively, a relative threshold could be used, but we found in a pilot study that our pre-computation works better in terms of average precision. We discuss the effect of the support threshold in \secref{sec:discussion}.
Then, we run the frequent subgraph miner for the threshold found via the approximate tree miner. Step 3 of \figref{fig:approach} shows this for our running example. We start with a set of connected components and the graph miner returns a set of frequent subgraphs, namely $\{g_1,g_2,g_3\}$ with $g_1 \subset g_2 \subset g_3$.
We use \textsc{Gaston}~\cite{Nijssen2005} graph miner, since it performed best (in terms of runtime) among the miners that we experimented with (\textsc{gSpan}, \textsc{Gaston} and \textsc{DIMSpan}) in a pilot study. \changed{In our pilot study, we ran the miners on a small selection of our datasets and experimented with the parameters of the miners. For many datasets, \textsc{gSpan} and \textsc{DIMSpan} did not terminate at all (we canceled the execution after 48h). \textsc{Gaston} (with embedding lists) was able to terminate in less then 10s on most of our datasets but consumes a lot of memory, typically between 10GB-25GB, which was not a problem for our 32GB machine in the pilot study. To rule out any effects due to approximate mining, we considered only exact miners. Therefore, we also could not use \textsc{Subdue} \cite{Ketkar2005}, which directly tries to optimize compression. Furthermore, \textsc{Subdue} was not able to discover both \eo{s} in the second experiment (see \secref{sec:evaluation}), without iterative mining and allowing for overlaps. Enabling these two options, \textsc{Subdue} did not terminate on more than 75\% of the pilot study datasets}. 
For frequent subtree mining, we use \textsc{Hops} \cite{Welke2020} because it provides low error rates and good runtime guarantees.\par
\textbf{Step 4: Select the most relevant subgraphs}:
Motivated by the minimum description length principle, \changed{which has been successfully applied to many different kinds of data}~\cite{grunwald2007minimum}, the most relevant patterns should not be the most frequent ones but the ones that give us a maximum compression for our original data \cite{Djoko1994}. 
That is, we want to express the given SCGs by a set of subgraphs with the property that the description length for the subgraphs together with the length of the description of the SCGs in terms of the subgraphs becomes minimal. \changed{This can be understood by looking at the corner cases. A single change has a large frequency but is typically not interesting. The entire model difference is large in terms of changes but has a frequency of only one and is typically also not an interesting \eo{}. ``Typical \eo{s}'' are therefore somewhere in the middle. We will use our experiments in \secref{sec:evaluation} to validate whether this assumption holds.
We define the compression value by}
$\operatorname{compr}(g) = \big({\operatorname{supp}}(g)-1\big) \cdot \big(|V_g| + |E_g|\big),$ 
where ${\operatorname{supp}}(g)$ is the support of $g$ in our set of input graphs (i.e., the number of components in which the subgraph is contained). The ``$-1$'' in the definition of the compression value comes from the intuition that we need to store the definition of the subgraph, in order to decompress the data again. \changed{The goal of this step is to detect the subgraphs from the previous step with a high compression value.}
The subgraphs are organized in a \emph{subgraph lattice}, where each graph has pointers to its direct subgraphs. Most of the subgraph miners already compute a subgraph lattice, so we do not need a subgraph isomorphism test here. Due to the downward closure property of the support, all subgraphs of a given (sub-)graph have, at least, the same frequency (in transaction-based mining). When sorting the output, we need to take this into account, since we are only interested in the largest possible subgraphs for some frequency. We therefore prune the subgraph lattice. 
The resulting list of recommendations is then sorted according to the compression value. \changed{Other outputs are conceivable, but in terms of evaluation, a sorted list is typical for a recommender system~\cite{schroder2011setting}.}  \par
More technically, let $SG$ be the set of subgraphs obtained from step 3. We then remove all the graphs in the set 
\begin{equation*}
\begin{split}
SG^- = \big\lbrace g &\in SG \,|\, \exists \tilde g \in SG \text{, with } g \subseteq \tilde g \\
&\wedge \operatorname{supp}(g) = \operatorname{supp}(\tilde g) 
 \wedge \operatorname{compr}(g) \leq \operatorname{compr}(\tilde g) \big \rbrace.
\end{split}
\end{equation*}
Our list of recommendations is then $SG \setminus SG^-$, sorted according to the compression metric.




For our running example in step 4 of \figref{fig:approach}, assume that the largest subgraph $g_3$ occurs 15 times (without overlaps). Even though the smaller subgraph $g_1$ occurs twice as often, we find that $g_3$ provides the best compression value and is therefore ranked first. The subgraph $g_2$ will be pruned, since it has the same support as its supergraph $g_3$ but a lower compression value.
We implement the compression computation and pruning using the \emph{NetworkX} Python library.


\textbf{Step 5: Generate \eo{s}}:
As a result of step 4, we have an ordered list of ``relevant'' subgraphs of the SCGs. We need to transform these subgraphs into model transformations that specify our learned \eo{s}. As shown in step~5 of \figref{fig:approach}, the subgraphs can be transformed to Henshin transformation rules in a straightforward manner. \changed{We use \textsc{Henshin} because it is used for the semantic lifting approach in our case study from Sec. \ref{sec:case-study}. In principle, any transformation language that allows us to express endogenous, in-place model transformations could be used instead. A survey of model transformation tools is given by Kahani et al.~\cite{kahani2019survey}.}




\section{Evaluation}\label{sec:evaluation}
\subsection{Research Questions}
We evaluate \textsc{Ockham} w.r.t.\ the following research questions:
\begin{itemize}
     
\item \begin{bfseries} RQ 1: \end{bfseries}  \textit{Is the approach able to identify \eo{s} that have actually been applied in model repositories?} If we apply some operations to models, the approach should be able to discover these from the data. Furthermore, when different \eo{s} are applied and overlap, it should be possible to discover them.
    
\item \begin{bfseries} RQ 2: \end{bfseries} \textit{Is the approach able to find typical \eo{s} or editing scenarios in a real-world setting?} Compared to the first research question, the approach should also be able to find typical scenarios in practice when we do not know which operations have been actually applied to the data. Furthermore, it should be possible to derive these \eo{s} in a real-world setting with large models and complex meta-models.
    
\item \begin{bfseries} RQ 3: \end{bfseries} \textit{What are the main drivers for the approach to work or fail?} We want to identify the characteristics of the input data or parameters having a major influence on the approach.
    
\item \begin{bfseries} RQ 4: \end{bfseries} \textit{What are the main parameters for the performance of the frequent subgraph mining?} Frequent subgraph mining has a very high computational complexity for general cyclic graphs. We want to identify the characteristics of the data that influence the mining time.
\end{itemize}

\changed{For RQ 1, we want to rediscover the \eo{s} from our ground truth, whereas in RQ 2, the discovered operations could also be some changes that are not applied in ``only one step'' but appear to be typical for a domain expert. We refer to the actually applied \eo{s} and the ones considered as typical by a domain expert as ``meaningful''.}
\subsection{Experiments}
We conduct three experiments to evaluate our approach. In the first two experiments, we run the algorithm on \changed{synthetic model repositories}. We know the ``relevant \eo{s}'', since we define them, and apply them to sample models. We can therefore use these experiments to answer RQ 1.
Furthermore, since we can control many properties of our input data for these simulated repositories, we can also use them to answer RQ3 and RQ4. In the third experiment, we apply \textsc{Ockham} to the dataset from our case study presented in \secref{sec:case-study} to answer RQ 2. \changed{The first two experiments help us to find the model properties and the parameters the approach is sensible to. Their purpose is to increase the internal validity of our evaluation. In addition, to increase external validity, we apply the approach in a real-world setting. None of the experiments alone can provide sufficient internal or external validity~\cite{siegmund2015views} but the combination of all experiments is suitable to assess whether \textsc{Ockham} can discover relevant \eo{s}.} 


\textbf{Experiment 1:}
As a first experiment, we simulate the application of \eo{s} on a simple component model. The meta-model is shown in \figref{fig:difference-graph}.\par
\textit{Setup:} 
For this experiment, we only apply one kind of \eo{} (the one from our running example in \figref{fig:approach}) to a random model instance. The Henshin rule specifying the operation consists of a graph pattern comprising 7 nodes and 7 edges. We create the model differences as follows:
We start with an instance $m_0$ of the simple component meta-model with 87 Packages, 85 Components, 85 SwImplementations, 172 Ports, 86 Connectors and 171 Requirements.
Then, the \eo{} is randomly applied $e$ times to the model 
to obtain a new model revision $m_1$. This procedure is then applied iteratively $d$ times to obtain the ``model history''
   $m_0 \xrightarrow[]{} m_1 \xrightarrow[]{} \dots m_{d - 1} \xrightarrow[]{} m_{d}.$
Each evolution step $m_i \xrightarrow[]{} m_{i+1}$ yields a difference $\Delta(m_i,m_{i+1})$.
To each application of the \eo{}, we apply a random perturbation. More concretely, a perturbation is another \eo{} that we apply with a certain probability $p$. This perturbation is applied such that it overlaps with the application of the main \eo{}. 
We use the tool \textsc{Henshin} \cite{Biermann2012} to apply model transformations to one model revision.
We then build the difference of two successive models as outlined in \secref{sec:appraoch}. 
In our experiment, we control the following parameters for the generated data. 

\begin{itemize}
    \item $d$: The number of differences in each simulated model repository. For this experiment, $d \,\in\, \lbrace10,20 \rbrace$.
    \item $e$: The number of \eo{s} to be applied per model revision in the repository, that is, how often the edit operation will be applied to the model. For this experiment, $e \,\in\, \lbrace 1, \dots , 100\rbrace$.
    \item $p$: The probability that the operation will be perturbed. For this experiment, we use $p \in \lbrace 0.1, 0.2, \dots, 1.0 \rbrace$.
\end{itemize}

This gives us 2000 (= 2x100x10) datasets for this experiment. A characteristics of our datasets is that increasing $e$, the probability of changes to overlap increases. Eventually, adding more changes even decreases the number of components in the SCG while increasing the average size of the components. 

Our algorithm suggests a ranking of the top $k$ subgraphs (which eventually yield the learned edit operations). In the ranked suggestions of the algorithm, we then look for the position of the ``relevant \eo{}'' by using a graph isomorphism test. To evaluate the ranking, we use the ``mean average precision at k'' (MAP@k) which is commonly used as an accuracy metric for recommender systems~\cite{schroder2011setting}:
\begin{equation*}
\operatorname{MAP@k} := \frac{1}{|D|} \sum_{\text{ds} \in \text{D}} \operatorname{AP@k} \, \text{,}
\end{equation*}
where $D$ is the family of all datasets (one dataset represents one repository) and AP@k is defined by 
\begin{equation*}
\operatorname{AP@k} := \frac{\sum_{i=1}^{k} \operatorname{P}(i) \cdot \operatorname{rel}(i)}{{|\text{total relevant subgraphs}|}} \, \text{, } 
\end{equation*}
where P($i$) is the precision at $i$, and rel($i$) indicates if the graph at rank $i$ is relevant. 
For this experiment, the number of relevant \eo{s} (or subgraphs to be more precise) is always one. Therefore, we are interested in the rank of the correct \eo{}. Except for the case that the relevant \eo{} does not show up at all, MAP@$\infty$ gives us the mean reciprocal rank and therefore serves as a good metric for that purpose.

For comparison only, we also compute the MAP@k scores for the rank of the correct \eo{s} according to the frequency of the subgraphs. Furthermore, we investigate how the performance of the subgraph mining depends on other parameters of \textsc{Ockham}. We are also interested in how average precision (AP), that is, AP@$\infty$, depends on the characteristics of the datasets. Note that for the first two experiments, we do not execute the last canonical step of our approach (i.e., deriving the \eo{} from a SCG), but we directly evaluate the resulting subgraph from step 4 against the SCG corresponding to the \eo{}. We run the experiments on an Intel® Core™ i7-5820K CPU @ 3.30GHz × 12 from which we use 3 cores per dataset and 31.3 GiB RAM.

To evaluate the performance of the frequent subgraph miner on our datasets, we fixed the relative threshold (i.e., the support threshold divided by the number of components in the graph database) to $0.4$. We re-run the algorithm for this fixed relative support threshold and $p \leq 0.4$. 

\par


\begin{table*}
\parbox{.45\linewidth}{
\caption{The MAP@k scores for the results using compression and frequency for the first experiment.}
\begin{adjustbox}{center}
\begin{tabular}{r|c c c c}
\toprule
   & \textbf{MAP@1} & \textbf{MAP@5} & \textbf{MAP@10} & \textbf{MAP@$\infty$}  \\
   \midrule
     \textbf{Compression} & \cellcolor[gray]{0.9}0.967 & \cellcolor[gray]{0.9}0.974 & \cellcolor[gray]{0.9}0.975 & \cellcolor[gray]{0.9}0.975   \\
      \textbf{Frequency}   & 0.016 & 0.353 & 0.368 & 0.368   \\
      \bottomrule
    \end{tabular}
    \end{adjustbox}
\label{tbl:exp1_mapatk}
}
\hfill
\parbox{.48\linewidth}{
\caption{The MAP@k scores for the results using compression and frequency for the second experiment.}
\begin{adjustbox}{center}
\begin{tabular}{r|c c c c}
\toprule
   & \textbf{MAP@2} & \textbf{MAP@5} & \textbf{MAP@10} & \textbf{MAP@$\infty$}  \\
   \midrule
     \textbf{Compression} & \cellcolor[gray]{0.9}0.955 &  \cellcolor[gray]{0.9}0.969 &  \cellcolor[gray]{0.9}0.969  &  \cellcolor[gray]{0.9}0.969   \\
      \textbf{Frequency}   & 0.013 & 0.127 & 0.152 & 0.190  \\
      \bottomrule
    \end{tabular}
   \end{adjustbox}
\label{tbl:exp2_mapatk}
}
\end{table*}

\textit{Results:} 
See Table \ref{tbl:exp1_mapatk} for the MAP@k scores for all datasets in the experiment. 
Table \ref{tbl:exp1_corr} shows the spearman correlation of the independent and dependent variables. If we look only on datasets with a large number of applied \eo{s}, $e > 80$, the spearman correlation for average precision vs. $d$ and average precision vs. $p$ becomes 0.25 (instead of 0.12) and -0.14 (instead of -0.07), respectively.
The mean time for running \textsc{Gaston} for our datasets was 1.17s per dataset.

\textit{Observations:}
We observe that increasing the number of \eo{s} has a negative effect on the average precision. Increasing the perturbation has a slightly negative effect, which becomes stronger for a high number of applied \eo{s} and therefore 
 when huge connected components start to form. The number of differences $d$ (i.e., having more examples) has a positive effect on the rank, which is rather intuitive.
We also observe a strong spearman correlation of the mining time with the number of applied \eo{s} $e$ (0.89) and implicitly also the average number of nodes per component (0.83). If we only look at the \eo{s} with rank $> 1$, we can also see a strong negative correlation of $-0.51$ with the average precision (not shown in Table \ref{tbl:exp1_corr}). This actually means that large mining times usually come with a bad ranking. 

\begin{table*}[h!]
\parbox{.45\linewidth}{
\caption{Spearman correlations for the first experiment.}
    \begin{adjustbox}{center}
    \begin{tabular}{r|c c c c c}
    \toprule
    & \textbf{p}  & \textbf{Mining} & \textbf{e}  & \textbf{d}  & \textbf{$\varnothing $ \#Nodes} \\
    & & \textbf{Time} &  &  &\textbf{per Comp} \\
    \midrule
    \textbf{AP} & \cellcolor[gray]{0.95}-0.07 & \cellcolor[gray]{0.97}-0.24 & \cellcolor[gray]{0.97}-0.23 & \cellcolor[gray]{0.75}0.12 &  \cellcolor[gray]{0.96}-0.21      \\
    \textbf{AP (for $\mathbf{e>80}$)} & \cellcolor[gray]{0.97}-0.14 & \cellcolor[gray]{0.95}-0.19 & \cellcolor[gray]{0.95}-0.19 & \cellcolor[gray]{0.70}0.25 &  \cellcolor[gray]{0.88}-0.03      \\
    \textbf{Mining Time} & \cellcolor[gray]{0.80}0.12 & - & \cellcolor[gray]{0.60}0.89 & \cellcolor[gray]{0.68}0.26 &  \cellcolor[gray]{0.62}$\,$0.83      \\
    \bottomrule
    \end{tabular}
    \end{adjustbox}
			\label{tbl:exp1_corr}
}
\hfill
\parbox{.5\linewidth}{
\caption{The Spearman correlation matrix for the second experiment.}
    \begin{adjustbox}{center}
    \begin{tabular}{r|c c c c c}
    \toprule
    & \textbf{p} & \textbf{Size at} & \textbf{Mining} & \textbf{e} & \textbf{$\varnothing $ \#Nodes} \\
    & & \textbf{Threshold} & \textbf{Time} &  & \textbf{per Comp} \\
    \midrule
    \textbf{AP} & \cellcolor[gray]{0.95}-0.31 & \cellcolor[gray]{0.85}-0.05 & \cellcolor[gray]{0.95}-0.25 & \cellcolor[gray]{0.88}-0.07 & \cellcolor[gray]{0.9}-0.19      \\
    \textbf{p} & - & \cellcolor[gray]{0.78}0.20 & \cellcolor[gray]{0.78}0.27 & \cellcolor[gray]{0.85}0 & \cellcolor[gray]{0.78}0.30   \\
    \textbf{Size at Threshold}  & - & - & \cellcolor[gray]{0.78}0.53 & \cellcolor[gray]{0.78}0.51 & \cellcolor[gray]{0.78}0.58    \\
    \textbf{Mining Time} & - & - & - & \cellcolor[gray]{0.7}0.87 & \cellcolor[gray]{0.7}0.92   \\
    \textbf{e} & - & - & - & - & \cellcolor[gray]{0.7}0.92    \\
    \bottomrule
    \end{tabular}
    \end{adjustbox}
\label{tbl:exp2_corr}
}
\end{table*}

\textbf{Experiment 2:}
In contrast to the first experiment, we want to identify more than one \eo{} in a model repository. We therefore extend the first experiment by adding another \eo{} and apply each of the operations with the same probability. In order to test if \textsc{Ockham} also detects \eo{s} with smaller compression than the dominant (in terms of compression) \eo{}, we choose the second operation to be smaller, its Henshin rule graph pattern comprises 4 nodes and 5 edges. \changed{It corresponds to adding a new \textit{Component} with its \textit{SwImplementation} and a \textit{Requirement} to a \textit{Package}.}\par 

\textit{Setup:}
Since the simulation of model revisions currently consumes a lot of compute resources, we fixed $d=10$ and considered only $e<=80$ for this experiment. The rest of the experiment is analogous to the first experiment.

\textit{Results:}
In Table \ref{tbl:exp2_mapatk} we give the MAP@k scores for this experiment. Table \ref{tbl:exp2_corr} shows the correlation matrix for the second experiment.

\textit{Observations:}
We can see that our compression-based approach clearly outperforms the frequency-based approach used as a baseline. From Table \ref{tbl:exp2_corr}, we can observe a strong dependency of the average precision on the perturbation parameter and the mining time.

\textbf{Experiment 3:}
Of course, the power of the simulation to mimic a real-world model evolution is limited. Especially, the assumption of random and independent applications of \eo{s} is questionable.
Therefore, for the third experiment, we use a real-world model repository from the railway software development domain (see \secref{sec:case-study}). Here, we do not know the operations that have actually been applied. We therefore compare the mined \eo{s} with \eo{s} randomly generated from the meta-model, and want to show that the mined \eo{s} are significantly more ``meaningful'' than the random ones. We will use the results from this interview to answer RQ2. \par
\textit{Setup:}
For this experiment, we mined
546 pairwise differences, with 4109 changes on average, which also contain changed attribute values (one reason for that many changes is that the engineering language has changed from German to English). The typical model size in terms of their abstract syntax graphs is 12081 nodes and, on average, 50 out of 83 meta-model classes are used as node types. 

To evaluate the quality of our recommendations, we conducted a semi-structured interview with five domain experts of our industry partner: 2 system engineers working with one of the models, 1 system engineer working cross-cutting, 1 chief system architect responsible for the product line approach and the head of the tool development team. We presented them 25 of our mined \eo{s} together with 25 \eo{s} that were randomly generated out of the meta-model. The \eo{s} were presented in the visual transformation language of \textsc{Henshin} which we introduced to our participants. Using a 5-point Likert scale, \changed{we asked whether} the \eo{} represents a typical edit scenario (5), can make sense but is not typical (3), and does not make sense at all (1). We compare the means of the Likert score for the population of random \eo{s} and mined \eo{s} \changed{to determine whether the mined operations are typical or meaningful}. \par

\textbf{Null hypothesis $\mathbf{H_0}$:} \textit{The mined \eo{s} do not present a more typical edit scenario than random \eo{s} on average.}\par

We set the significance level to $\alpha = 0.01$.
If we can reject the null hypothesis, we conclude that the mined \eo{s} more likely present typical edit scenarios than the random ones.
In addition, we discussed the mined \eo{s} with the engineers that have not been considered to be typical. 


\textit{Results:}
\changed{We found some operations that are typical to the modeling language SysML, for example, one which is similar to the simplified operation in Figure \ref{fig:approach}. We also found more interesting operations, for example, the addition of ports with domain specific port properties. Furthermore, we were able to detect some rather trivial changes. For example, we can see that typically more than just one swimlane is added to an activity, if any. We also found simple refactorings, such as renaming a package (which also leads to changing the fully qualified name of all contained elements) or also some refactorings that correspond to conventions that have been changed, for example, activities were owned by so called ``system use cases'' before but have been moved into ``packages''.}
Table \ref{tbl:exp3-stats} shows the results for the Likert values for the mined and random \eo{s} for the five participants of our study. We can see that for all participants, the mean Likert score for the mined operations is significantly higher than the mean for the random operations. After their rating, when we confronted the engineers with the true results, they stated that the \eo{s} obtained by \textsc{Ockham} represent typical edit scenarios. According to one of the engineers, some of the \eo{s} ``can be slightly extended'' (see also \secref{sec:discussion}). Some of the \eo{s} found by \textsc{Ockham} but not recognized by the participants where identified ``to be a one-off refactoring that has been performed some time ago''. 

\begin{table}[htbp]
\caption{Statistics for the Likert values of the mined and random \eo{s}.}
\begin{adjustbox}{center}
\begin{tabular}{$c^c^c^l^l}
\toprule
\rowstyle{\bfseries} Participant & mean & mean & p-value & p-value\\
\rowstyle{\bfseries} & mined & random & (t-test) & (Wilcoxon)\\
\midrule
P1          & 3.20       & 1.68        & $11.8\cdot10^{-5}$  & $29.0\cdot10^{-5}$  \\
P2          & 4.04       & 2.76        & $16.6\cdot10^{-4}$ & $6.43\cdot10^{-3}$ \\
P3          & 4.32       & 2.60        & $9.30\cdot10^{-6}$ & $5.87\cdot10^{-5}$  \\
P4          & 4.32       & 1.08        & $2.67\cdot10^{-15}$ & $3.51\cdot10^{-10}$ \\
P5          & 4.48       & 1.60        & $1.17\cdot10^{-11}$ & $1.15\cdot10^{-7}$ \\
\midrule
\textbf{Total} & \textbf{4.072} &\textbf{1.944}  & $\mathbf{<2.2\cdot10^{-16}}$ &  $\mathbf{<2.2\cdot10^{-16}}$\\
\bottomrule
\end{tabular}
\end{adjustbox}
\label{tbl:exp3-stats}
\end{table}

\textit{Observations:}
The \eo{s} found by \textsc{Ockham} obtained significantly higher (mean) Likert scores than the random \eo{s}. We can therefore reject the null hypothesis and conclude that, compared to random ones, our mined \eo{s} can be considered as typical edit scenarios on average. Furthermore a mean Likert score of almost 4.1 shows that the \eo{s} are considered as typical on average. 
In \secref{sec:discussion}, we take a closer look at the \eo{s} that where not considered as typical edit scenario by the participants.
\subsection{Discussion}\label{sec:discussion}
In the first two experiments, we can see from the high MAP@k values that \textsc{Ockham} is able to recover the \eo{s} that have been applied. Furthermore, the third experiment shows that \textsc{Ockham} provides meaningful \eo{s} in a real-world setting. The observations from the first experiment suggests that the main driver for the performance of the frequent subgraph mining is the average number of nodes of our SCGs and the number of \eo{s} applied in the evolution steps yielding our model differences. To answer RQ3, we have to take a closer look at the datasets for which our approach gives non-optimal results.

\subsubsection{Reasons for non-optimal results}\label{sec:non_optimal}
We have to distinguish between the two cases that (1) the correct \eo{} is not detected at all and (2) the correct \eo{} has a low rank, i.e., appears later in the ranked list.

\textit{Edit operation has not been detected:}
For the second experiment, in 22 out of 800 examples, \textsc{Ockham} was not able to detect both edit operations. In 10 of these cases the threshold has been set too high. To mitigate this problem, in the real-world setting, the threshold parameters could be manually adjusted until the results are more plausible. In the automatic approach, further metrics have to be integrated. 
Other factors that cause finding the correct \eo{s} to fail are the perturbation, average size of component and the size of the component ``at threshold'', as can be seen from Table \ref{tbl:exp1_na_suspects}.\par

\begin{table}[]
\caption{The main drivers for \textsc{Ockham} to fail in detecting the correct subgraph in experiment 1.}
\begin{adjustbox}{center}
    \begin{tabular}{l|r r r r}
    \toprule
         &  &  \textbf{Average Size of} & &\\
         & \textbf{p} & \textbf{a Component} & \textbf{Size at} & \textbf{Mining} \\
         & & \textbf{(\# of Nodes)} & \textbf{Threshold} & \textbf{Time} \\
    \midrule
   \textbf{Overall Mean} & 0.55 & 57.6 & 8.20 & 1.26 \\
    \begin{tabular}[l]{@{}l@{}} \textbf{Mean for un-} \\ \textbf{detected operation}\end{tabular} & 0.79 & 109.0 & 10.03 & 2.55\\
    
    \bottomrule
    
    \end{tabular}
    \end{adjustbox}
\label{tbl:exp1_na_suspects}
\vspace{-1em}
\end{table}

Given a support threshold $t$, the \emph{size at threshold} is the number of nodes of the $t$-largest component. The intuition behind this metric is the following: For the frequent subgraph miner, in order to prune the search space, a subgraph is only allowed to appear in, at most, $t-1$ components. Therefore, the subgraph miner needs to search for a subgraph, at least, in one component with size greater than the \emph{size at threshold}.  Usually, the size of a component plays a major role in the complexity of the subgraph mining. When the $t$-largest component is small, we could always use this component (or smaller ones) to guide the search through the search space and therefore we will not have a large search space. So, a large size of the component at threshold could be an indicator for a complicated dataset.\par
We clearly see that perturbation, average size of a component, and the size at threshold are increased for the datasets for which our approach does not perform well. 
We looked deeper into the results of the datasets from the first experiment for which the correct subgraph has not been identified. We can see that, for some of these subgraphs, there is a supergraph in our recommendations that is top ranked. Usually this supergraph contains one or two additional nodes. Since we have a rather small meta-model and we only use four other \eo{s} for the perturbation, it can happen rarely, that these larger graphs occur with the same frequency as the actual subgraph. The correct subgraphs are then pruned.

\textit{Edit operation has a low rank:}
First, note that we observe a low rank (rank $\geq 5$) only very rarely. For the first experiment, it happened in 7 out of 2000 datasets, while for the second experiment, it did not happen at all. In Table \ref{tbl:exp1_low_rank_drivers}, we list the corresponding datasets and the values for drivers of a low rank. 
\begin{table}[]
\caption{Possible drivers for a low rank ($\geq 5$).}
\label{tbl:exp1_low_rank_drivers}
\begin{adjustbox}{center}
    \begin{tabular}{r r r r r l r}
    \toprule
         &  &  & \textbf{Average} &  &  \\
       \textbf{d}  & \textbf{e} & \textbf{p} & \textbf{\#Nodes per} & \textbf{Size at} & \textbf{Average} & \textbf{Rank} \\
        &  &  &  \textbf{Component} & \textbf{Threshold} & \textbf{Precision} & \\
    \midrule
    10 & 92 & 0.3 & 142.2 & 13 & 0.13 & 8 \\
    10 & 67 & 0.4 & 91.0 & 16 & 0.14 & 7 \\
    10 & 78 & 0.8 & 87.3 & 14 & 0.14 & 7 \\
    10 & 98 & 0.8 & 127.7 & 14 & 0.067 & 15 \\
    20 & 81 & 0.1 & 227.0 & 16 & 0.13 & 8 \\
    20 & 99 & 0.1 & 272.2 & 19 & 0.010 & 99 \\
    20 & 100 & 0.1 & 272.7 & 17 & 0.013 & 78\\
 \bottomrule

    \end{tabular}
    \end{adjustbox}
\end{table}
One interesting observation is that, for some of the datasets with low ranked correct subgraph, we can see that the correct graph appears very early in the subgraph lattice, for example, first child of the best compressing subgraph but rank 99 in the output, or first child of the second best subgraph but rank 15 in the output. This suggests that this is more a presentation issue which is due to the fact that we have to select a linear order of all subgraph candidates for the experiment.

\subsubsection{Qualitative results}
We only found two mined \eo{s} that received an average Likert score below 3 from the five practitioners in the interviews.
The first one was a refactoring that was actually performed but that targeted only a minority of all models. Only two of the participants where aware of this refactoring and one of them did not directly recognize it due to the abstract presentation of the refactoring. The other \eo{} that was also not considered as a typical edit scenario was adding a kind of document to another document. This \eo{} was even considered as illegal by 3 out of the 5 participants. The reason for this is the internal modeling of the relationship between the documents, which the participants were not aware of. So, it can also be attributed to the presentation of the results in terms of Henshin rules, which require an understanding of the underlying modeling language's meta-model.\par
For four of the \eo{s}, some of the participants mentioned that the \eo{} can be extended slightly. We took a closer look at why \textsc{Ockham} was not able to detect the extended \eo{}, and it turned out that it was due to our simplifications of the locality relaxation and also due to the missing type hierarchies in our graphs. For example, in one \eo{}, one could see that the fully qualified name (name + location in the containment hierarchy) of some nodes has been changed, but the actual change causing this name change was not visible, because it was a renaming of a package a few levels higher in the containment hierarchy that was not directly linked to our change. Another example was a ``cut off'' referenced element in an \eo{}. The reason why this has been cut off was that the element appeared as several different sub-classes in the model differences and each single change alone was not frequent.\par
When looking at the mined \eo{s} it became clear, that the approach was able to implicitly identify constraints which where not made explicit in the meta-model. 

\subsubsection{Result summary}

    \begin{bfseries} RQ 1: Is this approach able to identify relevant \eo{s} in model repositories? \end{bfseries}
    We can answer this question with a ``yes''. Experiment 1 and 2 show high MAP scores. Only for a large number of applied operations and 
    a large size of the input graphs, the approach fails in finding the applied \eo{s}. 
    
    \begin{bfseries} RQ 2: Is this approach able to find typical edit operations or editing scenarios in a real-world setting?\end{bfseries}
    We could show that the approach is able to detect typical edit scenarios. The approach is therefore \emph{sound} to a large extend, and incomplete \eo{s} can be adjusted manually. We cannot state yet that the approach is also \emph{complete} (i.e., is able to find all relevant edit scenarios), though.
    
    \begin{bfseries} RQ 3: What are the main drivers for the approach to work or fail?\end{bfseries}
    The main drivers for the approach to fail are a large average size of a component and the size of the component at threshold (see definition in \secref{sec:non_optimal}). The average size is related to the number of \eo{s} applied per model difference. In a practical scenario, huge differences can be excluded when running \eo{} detection. The size of the component at threshold can, of course, be reduced by increasing the support threshold parameters of the frequent subgraph mining. With higher threshold, we increase the risk of missing some less frequent \eo{s} but the reliability for detecting the correct (more frequent) operations is increased. Having more examples improves the results of our approach.
    
    \begin{bfseries} RQ 4: What are the main parameters for the performance of the frequent subgraph mining?\end{bfseries}
    The main driver for the performance of the frequent subgraph mining is the number of applied \eo{s} per difference, which is related to the average number of nodes per component. Furthermore, we have a strong dependence between the average precision and the time spent for the frequent subgraph mining.

\section{Limitations and Threats to Validity}
\subsection{Limitations} \label{sec:limitations}
\textit{Locality relaxation:}
One limitation of our approach is the locality relaxation, which limits our ability to find patterns that are scattered across more than one connected component of the SCG. As we have seen in our railway case study, this usually leads to incomplete \eo{s}.
Another typical example for violating the relaxation are naming conventions.
In the future, we plan to use natural language processing techniques like semantic matching to augment the models by further references. 
\par
\textit{No attribute information:}
For this study, we did not take attribute information into account. Attributes (e.g., the name of a component) could also be integrated into the \eo{} as preconditions or to extract the parameters of an \eo{}. 
For the purpose of summarizing a model difference or identifying violations in a model difference, preconditions and parameters are not important, though, but only the presence of structural patterns. 

\textit{Application to simplified graphs:}
An \eo{} generally is a model transformation. Model transformation engines such as \textsc{Henshin} provide also features to deal with class inheritance or multi-object structures (roughly speaking, for each expressions in model transformations). In our approach, we are not handling these features yet. 
They can be integrated into the approach in a post-processing step. 
For example, one possibility would be to feed the example instances of patterns discovered by \textsc{Ockham} into a traditional MTBE approach~\cite{Kehrer2017}. 
\par
\textit{Transient effects:}
We also do not take so-called transient effects into account yet. One applied \eo{} can invalidate the pre- or post-conditions of another \eo{}. 
However, we have seen in our experiments that it only causes problems in cases where we apply only a few ``correct'' \eo{s} with high perturbation. In the practical scenario, the ``perturbations'' will more likely cancel each other out.
When a transient effect occurs very frequently, a new pattern will be discovered. That is, when two (or more) operations are always applied together, we want to find the composite pattern and not the constituent ones.\par  
\textit{Focus on single subgraphs instead of sets:}
Another limitation is the fact that we focused the optimization on \emph{single} \eo{s} but not a \emph{complete set} of \eo{s}. 
One could detect only the most-compressing \eo{} and then substitute this in the model differences and re-run the mining to discover the second most-compressing \eo{} and so on. Another solution would be to detect a set of candidate \eo{s} using \textsc{Ockham} and then select an optimal set 
using a meta-heuristic search algorithm and optimizing the \emph{total compression}. 
We leave this for further research.\par



\subsection{Threats to validity}\label{sec:threats}
\textit{Internal validity:}
The first two experiments were designed so that we can control input parameters of interested and observe their effect on the outcome. \textsc{Ockham} makes assumptions such as the locality relaxation, which could risk the real-world applicability. Because of this and since we can not claim that the results from the first two experiments also hold true in a real-world setting, we additionally applied our approach to an industrial case study. We can therefore be confident that \textsc{Ockham} also gives reasonable results in a practical scenario.
In our simulations, we applied the \eo{} randomly to a meta-model. To reduce the risk of observations that are only a result of this sampling, we created many example models. In the real-world setting, we compared the mined edit operations to random ones to rule out ``patternicity'' \cite{shermer2008patternicity} as an explanation for high Likert rankings. None of our participants reported problems in understanding \textsc{Henshin}'s visual notation, which gives us a high confidence regarding their judgements. 
\changed{The participants of the interviews in the third experiment were also involved in the project where the model history was taken from. There might be the risk that the interviewees have only discovered operations they have ``invented''. In any case, because of the huge project size and because 22 out of 25 of the edit operations were recognized as typical by more than one of the participants, this is unlikely.}\par
\textit{External validity:}
Some of the observations in our experiments could be due to the concrete set of \eo{s} in the example or even due to something in the meta-models. In the future, \textsc{Ockham} has to be tested for further meta-models to increase the external validity of our results. We have validated our approach in a real-world setting, which increases our confidence in its practicality, though. 
Since we have used an exact subgraph miner, we can be sure that the discovered \eo{} are independent of the subgraph mining algorithm. 
\par

\section{Related Work}

\changed{Several approaches have been proposed to (semi-)automatically learn model transformations in the field of Model Transformation By Example.
In the first systematic approach of MTBE, Varró \cite{Varro2006} proposes an iterative procedure which tries to derive {\em exogenous} (i.e., source and target meta-model are different) model transformations by examples. Appropriate examples need to be provided for the algorithm to work.
Many approaches to learning exogenous model transformations have been proposed until now. For example,  Berramla et al. \cite{berramla2020} use statistical machine translation and language models to derive the transformations. Or Baki and Sahraoui \cite{Baki2016} apply simulated annealing  to learn the operations. 
Regarding \emph{exogenous} transformations there is also an approach by Saada et al. \cite{Saada2014} which uses graph mining techniques to learn concepts which are then used to identify new transformation patterns. 
}

\changed{However, as already mentioned in the introduction, most closely related to our approach is the area of MTBE for {\em endogenous} model transformations.}
Compared to exogenous MTBE, there are only a few studies available for endogenous MTBE. Brosch et al.~\cite{Langer2009} present a tool called the \textsc{Operation Recorder}, which is a semi-automatic approach to derive model transformations by recording all transformation steps. A similar approach is presented by Yun et al. \cite{Gray2011}, who also infer complex model transformations from a demonstration. Alshanqiti et al. \cite{Jalali2012} learn transformation rules from a set of examples by generalizing over pre- and postcondition graphs. Their approach has been applied to the derivation of \eo{s}, including negative application conditions and multi-object patterns \cite{Kehrer2017}. 
Instead of learning a single operation, Mokaddem et al. \cite{Mokaddem2018} use a genetic algorithm to learn a set of refactoring rule pairs of examples before and after the application of refactoring. The creation of candidate transformations that conform to the meta-model is done by the use of a ``fragment type graph'', which allows them to grow candidate patterns which conform to the meta-model. 
Their algorithm optimizes a model modification and preservation score. Ghannem et al. \cite{Ghannem2018} also use a genetic algorithm (i.e., NSGA-II) to learn model refactorings from a set of ``bad designed'' and ``good designed'' models. Their approach  distinguishes between structural similarity and semantic similarity and tries to minimize structural and semantic similarity between the initial model and the bad designed models and to maximize the similarity between the initial and the well designed models.
\par


All of these approaches for learning endogenous model transformations are (semi-)supervised. Either a concrete example is given (which only contains the transformation to be learned) or a set of positive and negative examples is given. In the case of Mokaddem et al.'s genetic approach, it is assumed that all transformations that can be applied are actually applied to the source models. 
For the meta-model used in our real-world case study, we do not have any labeled data. In general, we are not aware of any completely unsupervised approach to learn endogenous model transformations. To reduce the search space, we make use of the evolution of the models in the model repository, though. We do not directly work on the models as in the approaches above but work on structural model differences. \par

\changed{
Furthermore, there is related work in the source code domain.
Regarding one of our motivations for mining \eo{s}, namely to simplify differences, there are several approaches in the source code domain \cite{Yu2011, martinez2013}. 
These approaches are more comparable to the approach of semantic lifting \cite{Kehrer2011}, to aggregate or filter model differences according to given patterns but they are not learning the patterns themselves.
There are also approaches to mine change patterns in source code. For example, Dagit et al. propose an approach based on the abstract syntax tree (AST)~\cite{Dagit2013}, and Nguyen et al. mine patterns based on a so called fine-grained program dependence graph~\cite{nguyen2019}. 
There is also some work that focuses on mining design patterns from source code~\cite{Oruc2016, balanyi2003mining, ferenc2005design, dong2009review}.
The idea behind these approaches, that is, learning (change) patterns from a version history, is comparable to ours. Other than these approaches, \textsc{Ockham} works on a kind of abstract syntax graph which already includes domain knowledge given by the meta-model. 
Furthermore, we do not use a similarity metric to detect change groups or frequent changes but use an (exact) subgraph mining approach instead. In model-driven engineering, one often has some kind of identifiers for the model elements, which makes the differencing more reliable and removes the need for similarity-based differencing methods. }

\section{Conclusion and Outlook}
We proposed an approach, \textsc{Ockham},  for automatically deriving \eo{s} specified as in-place model transformations from model repositories, based on the idea that a meaningful \eo{} will be one which provides a good compression for the model differences. \textsc{Ockham} uses frequent subgraph mining on labeled graph representation of model differences to discover frequent patterns in the model differences. The patterns are then filtered and ranked based on a compression metric to get a list of recommendations for meaningful \eo{s}. To the best of our knowledge, \textsc{Ockham} is the first approach for learning domain-specific edit operations in a fully unsupervised manner, i.e., without relying on any manual intervention or input from a developer or domain expert. \par
We have successfully evaluated \textsc{Ockham} on two case studies using synthetic ground-truth EMF models and on a large-scale real-world case study in the railway domain. We find that our approach is able to extract \eo{s} that have actually been applied from the model differences and also discovers meaningful \eo{s} in a real-world setting. Too large connected components in the differences is the main driver for the approach to fail in discovering actually applied \eo{s}. Performance mostly depends on the number of applied \eo{s} in a model difference.
Our approach can be applied to models of any Domain-Specific Modeling Language for which model histories are available. New effective \eo{s} that are performed by the users can be learned at runtime and recommendations can be made.\par
For our future research, we plan to extend \textsc{Ockham} by a meta-heuristic search to identify the optimal set of operations. Another alternative approach which we want to study in the future is to use a clustering algorithm and then feed the clusters into the frequent subgraph mining step of our approach. This will allow us also to deal with examples where the connected components of the difference graph are huge. \par

\def\bibfont{\footnotesize}
\bibliographystyle{plain}
\bibliography{bib}

\end{document}